# Open Questions in the Theory of Semifeasible Computation[*]

Piotr Faliszewski and Lane A. Hemaspaandra
Department of Computer Science
University of Rochester
Rochester, NY 14627 USA.

June 20, 2005

## Abstract

The study of semifeasible algorithms was initiated by Selman's work a quarter of century ago [Sel79,Sel81,Sel82]. Informally put, this research stream studies the power of those sets $L$ for which there is a deterministic (or in some cases, the function may belong to one of various nondeterministic function classes) polynomial-time function $f$ such that when at least one of $x$ and $y$ belongs to $L$, then $f(x,y) \in L \cap \{x,y\}$. The intuition here is that it is saying: "Regarding membership in $L$, if you put a gun to my head and forced me to bet on one of $x$ or $y$ as belonging to $L$, my money would be on $f(x,y)$."

In this article, we present a number of open problems from the theory of semifeasible algorithms. For each we present its background and review what partial results, if any, are known.

## 1 Introduction

The most fundamental class of semifeasible sets is P-sel, which Selman defined in the late 1970s as a complexity-theoretic analog of the semirecursive sets from recursive function theory [Joc68]. P-sel is the class of languages $L$ for which there is a (total, deterministic) polynomial-time function that given any two strings $x$ and $y$ has the property that if exactly one of them is in the set then it outputs the one that is in the set. However, it is not hard to see that this yields the same class as the following definition.

**Definition 1.1 ([Sel79])** *A set $L$ is* P*-selective exactly if there is a (total, deterministic) polynomial-time function $f$ such that, for all $x, y \in \Sigma^*$:*

1. $f(x,y) \in \{x,y\}$, *and*

2. $\{x,y\} \cap L \neq \emptyset \implies f(x,y) \in L$.

---

[*]Supported in part by grant NSF-CCF-0426761. This TR is a draft of the March 2006 *SIGACT News* Complexity Theory Column. We welcome any feedback that will help us improve the article.

*We say that the function $f$ is a* P*-selector function for $L$. The set of all* P*-selective sets is denoted by* P*-sel.*

In the 1990s this notion was generalized in Hemaspaandra et al. [HHN$^+$95,HNOS96b] to study the analogous classes obtained when the selector function is instead allowed to range over each of the four standard classes of polynomial-time nondeterministic functions [BLS84, BLS85], NPSV$_{\text{total}}$, NPSV, NPMV$_{\text{total}}$, NPMV. In this article we will employ just the most interesting of these four cases. Namely, the NPSV-selective sets and their close relatives.

**Definition 1.2** *Let $\Sigma$ be some alphabet and let $f$ be a function from $\Sigma^*$ to $2^{\Sigma^*}$. We say that $f$ is an* NPMV *function if there exists a nondeterministic polynomial-time Turing machine $M$ such that:*

- *On each computation path $M$ outputs a string if and only if it accepts on that path.*

- *The set $M(x) = \{y \mid y \text{ is output on some computation path of } M \text{ when run with input } x\}$ is equal to the set $f(x)$.*

*We say that the machine $M$ computes $f$.*

**Definition 1.3** *A function $f : \Sigma^* \to 2^{\Sigma^*}$ belongs to the class* NPSV *if and only if $f$ belongs to* NPMV *and it holds that, for every $x$, $||f(x)|| \leq 1$.*

Note that the above definition does not prevent machines computing NPSV functions from accepting on more than one path on some inputs. It merely says that in such situations the machine is obligated to output the same value on each of the accepting paths.

**Definition 1.4** *A set $L$ is said to be* NPSV*-selective if there is a function $f \in$ NPSV such that, for all $x, y \in \Sigma^*$,*

1. *$f(x) \in \{\emptyset, \{x\}, \{y\}\}$, and*

2. *$L \cap \{x, y\} \neq \emptyset \implies \emptyset \neq f(x) \subseteq L$.*

*In this case we say that $f$ is an* NPSV*-selector for $L$. Note that if neither $x$ nor $y$ belongs to $L$, it is legal for $f(x, y)$ to equal $\emptyset$. The set of all* NPSV*-selective sets is denoted by* NPSV*-sel.*

There are many motivations for studying the semifeasible sets. For space reasons, rather than presenting here the many motivations we refer readers to the coverage of these that appears in the preface of the recent book of Hemaspaandra and Torenvliet [HT03] about the semifeasible sets. However, just to mention quickly one motivation for each case, one may view the P-selective sets as capturing a notion of heuristically guided search. That is, we may imagine the selector function as ideally choosing the "better" option in some search (after all, in a heuristically driven search, we often can deal with choices one at a time, and our key question is, "What do I do *now*?"). And, as to the NPSV-selective sets, the best motivation for them comes from their actual usefulness: This notion was the critical one needed to show that NP lacks unique solutions unless the polynomial hierarchy collapses. We will discuss this in a bit more detail as part of Question 3.

We stress that we in no way claim that the questions here are "the most important" open questions about semifeasible computation. Such a judgment would clearly be a matter of individual taste. However, we do feel that these questions are important, interesting, and just plain attractive. We also mention that in general these questions are not new to this article, but have been stated in the cited research papers or have been highlighted in earlier survey and open question papers; for those wanting additional references, we suggest the monograph [HT03] and the various existing survey and open question articles that are in part or in full about selectivity or closely related notions [DHHT94,NT03,FH]. These survey papers point via their references to the vast assortment of original-literature papers.

## 2 Open Questions

### 2.1 Question 1: Do All P-Selective Sets Have Linear Advice? Question 2: Are All P-Selective Sets Associatively P-Selective?

These two questions are closely related so we will handle them jointly.

First, let us quickly define the additional notions needed. Let $A$ be a P-selective set.

- We say that $A$ is *commutatively* P-*selective* if there is a commutative function $f$ that is a P-selector function for $A$.

- We say $A$ is *associatively* P-*selective* if there is an associative function $f$ that is a P-selector function for $A$.

- We say that $A$ is *associatively, commutatively* P-*selective* if there is an associative, commutative function $f$ that is a P-selector function for $A$.

We also need the notion of Karp–Lipton advice classes. A Karp–Lipton advice class captures the notion of the class of sets that can be accepted by a given complexity class "helped" by some extra advice that depends just on the input's length.

**Definition 2.1 ([KL80])** *Let $\mathcal{C} \subseteq 2^{\Sigma^*}$ and let $\mathcal{F}$ be a class of functions (each mapping from $\mathbb{N}$ to $\mathbb{N}$).*

$$\mathcal{C}/\mathcal{F} = \{L \mid (\exists C \in \mathcal{C})(\exists f \in \mathcal{F})(\exists g : \mathbb{N} \to \Sigma^*)[(\forall n)[|g(n)| = f(n)] \wedge L = \{x \mid \langle x, g(|x|)\rangle \in C\}]\}.$$

When appearing in an advice class (e.g., P/linear), we will use linear, quadratic, and poly to refer respectively to the class of all linearly bounded, quadratically bounded, and polynomially bounded functions.

So, we can now turn to asking about the advice complexity of the P-selective sets. The first advice complexity result for the P-selective sets was the following lovely theorem due to Ko.

**Theorem 2.2 ([Ko83])** P-sel $\subseteq$ P/quadratic.

Ko's proof of this draws on the close connection between commutatively selective sets and tournaments, and also exploits the fact that all P-selective sets are commutatively P-selective.

Can Ko's quadratic-advice claim be replaced by a linear-advice claim? This was first achieved by Hemaspaandra et al., who showed that by increasing the strength of the advice "interpreter" from P to PP (probabilistic polynomial time), and by using a census/threshold argument, one could lower the advice from quadratic to linear.

**Theorem 2.3 (Implicit in [HNOS96a])** P-sel $\subseteq$ PP/linear.

That result was itself strengthened to allow even NP interpreters to achieve linear advice.

**Theorem 2.4 ([HT96], see also [HNP98])** P-sel $\subseteq$ NP/linear $\cap$ coNP/linear.

Theorem 2.4 is proved via a cute, half-century old fact from tournament theory: In any tournament, there will be at least one node that is a "king," i.e., that can reach all other nodes via paths of length at most two [Lan53].

The linear upper bound is tight down to the bit. P-sel $\subseteq$ NP/$n+1$ $\cap$ coNP/$n+1$, yet P-sel $\not\subseteq$ NP/$n$ and P-sel $\not\subseteq$ coNP/$n$. (In fact, P-sel $\not\subseteq$ EEEEEEE/$n$; the bound is less about complexity than about an information bottleneck [HT96]).

So, returning to Question 1, yes, all P-selective sets have linear (indeed $n+1$ bit) advice with respect to NP advice interpreters. However, do all P-selective sets have linear advice with respect to P interpreters? That is, does P-sel $\subseteq$ P/linear hold?

Currently, this is an open question. Indeed, it is not even known whether there exists any $\epsilon > 0$ such that P-sel $\subseteq$ P/$n^{2-\epsilon}$. However, some relevant results are known. Via a quite complex Kolmogorov-complexity based proof, Thakur [Tha03] has shown that there is an oracle $W$ such that there are $P^W$-selective sets that are not in $P^W$/linear. Thus, no relativizable proof can establish P-sel $\subseteq$ P/linear. Of course, proving (in the real world) that P-sel $\not\subseteq$ P/linear would be earthshaking, since in light of Theorem 2.4 this would as a side effect prove P $\neq$ NP.

The other line seeking to get some insight into whether P-sel $\subseteq$ P/linear involves the associatively P-selective sets. In particular, it is easy to see the following.

**Theorem 2.5 ([HHN04])** *Each set that is associatively, commutatively P-selective belongs to P/linear (and even to P/$n+1$).*

Also, one can show the following result.

**Theorem 2.6 ([HHN04])** *Each associatively P-selective set is associatively, commutatively P-selective.*

Thus, it is clear that "all P-selective sets are associatively P-selective" is a sufficient condition for "all P-selective sets belong to P/linear." This relation is the reason for discussing together the advice complexity of the P-selective sets and the algebraic properties of their P-selector functions.

It is not known if this sufficient condition is necessary. Interestingly enough, though, all currently known ways of constructing P-selective sets (in the real world) yield only associatively P-selective sets (and thus sets belonging to P/linear). On the other hand, Thakur's [Tha03] oracle mentioned earlier clearly—since Theorems 2.5 and 2.6 both relativize—shows that in some worlds there are P-selective sets that are not associatively P-selective.

A related interesting open issue is what consequences—in addition to P-sel $\subseteq$ P/linear—follow from assuming that all P-selective sets are associatively P-selective. Most particularly, do any complexity class collapses follow? Note that for one type of nondeterministic selectivity (that we will not define here), a result of this sort is known: If all NPMV-selective sets are associatively NPMV-selective, then PH = $S_2^{NP}$ [HHN04].

Finally, regarding Theorem 2.4, we mentioned that the half-century old notion of a king—a node that can reach all the other nodes via paths of length at most two—can help yield results about semifeasible computation. Indeed, when one has a type of node such that if the P-selective set is nonempty at a length then all nodes of that type belong to the P-selective set, then one can use the complexity of testing-for or finding that type of node to obtain nonimmunity results (see [HOZZ]). Kings have that property: If a P-selective set $A$ is nonempty at a certain length, then each king it has of that length (in the tournament at that length associated with some fixed commutative P-selector function for $A$) belongs to $A$. So, from this and the $\Pi_2^p$ upper bound for testing kingship in succinctly specified tournament families ([HOZZ], and recently the kingship problem has been shown to even be $\Pi_2^p$-complete [HHW05]), one may obtain nonimmunity results for the P-selective sets [HT05]. For example, all infinite P-selective sets have infinite $\Pi_2^p/1$ sets, and more interestingly, it holds that although P-sel is EXP-immune it is not EXP/1-immune (so its immunity structure is so fragile that it can be shattered by a single bit per length) [HT05]. And yet another use of kings in theoretical computer science is that kings have been used to study the first-order definability of certain reachability questions [NT02].

Readers particularly interested in Questions 1 and 2 may wish to look at the more detailed tutorial [FH], and at the references therein.

## 2.2 Question 3: How Far Does NP3V $\subseteq_c$ NP2V Collapse PH?

Before we describe the problem, its background, and relation to selectivity, let us define the notions involved.

Recall that NPSV is the class of functions computed by nondeterministic polynomial-time Turing machines that output at most one distinct value. In this section we will be concerned with NPSV's close cousins, the NP$k$V functions.

**Definition 2.7 ([NRRS98], see also [HOW02])** *Let $k \in \{1, 2, 3, \ldots\}$. We say that a function $f : \Sigma^* \to 2^{\Sigma^*}$ belongs to the class NP$k$V exactly if $f$ belongs to NPMV and it holds that, for every $x$, $||f(x)|| \leq k$.*

The class NP1V is our familiar class NPSV.

It is natural to ask about the relation between NP$k$V and NP$(k+1)$V. In particular, what is the relation between NPSV and NP2V? Of course, there is no point in asking whether these sets are equal, because they clearly are not; a function with two distinct values on some $x$ cannot be equal to a function with only one. Thus, to compare such function classes we use the notion of refinement. A function $g$ is a *refinement* [Sel94] of a function $f$ if it has the same domain and, for every $x$, $g(x) \subseteq f(x)$. To put it more formally, $g$ is a refinement of $f$ if and only if:

1. $g(x) \subseteq f(x)$, and
2. $f(x) \neq \emptyset \implies g(x) \neq \emptyset$.

Let $\mathcal{F}$ and $\mathcal{G}$ be two classes of functions. We say that $\mathcal{F}$ is refined by $\mathcal{G}$, denoted by $\mathcal{F} \subseteq_c \mathcal{G}$, if and only if every member of $\mathcal{F}$ has a refinement in $\mathcal{G}$ [Sel94].

Armed with the above definitions we may describe the line of research that aims to establish the connection between NP$k$V and NPSV functions. We are interested in determining whether every NP2V function has an NPSV refinement. The first result we discuss gives a strong answer to this question. In the mid-90s Hemaspaandra et al. [HNOS96b] showed that if all NPMV functions had NPSV refinements then the polynomial hierarchy would collapse to its second level. Their proof only requires all NP2V functions to have NPSV refinements, and actually yields a stronger collapse as we discuss further on.

**Theorem 2.8 ([HNOS96b])** *If* NP2V $\subseteq_c$ NPSV *then* PH $= \Sigma_2^p$.

A natural follow-up question is whether a similar collapse can be established by assuming that NP$k$V is refined by NP$(k-1)$V, for various $k > 2$. The answer was provided soon afterward by Naik et al. [NRRS98] in the following theorem.

**Theorem 2.9 ([NRRS98], see also [Ogi96, HOW02])** *For each integer* $k > 1$*, if* NP$k$V $\subseteq_c$ NP$(k-1)$V *then* PH $= \Sigma_2^p$.

These results were further improved by Hemaspaandra, Ogihara, and Wechsung [HOW02] who showed that Theorem 2.8 and Theorem 2.9 are a consequence of a more general lowness result. This was an improvement as if one believes that the polynomial hierarchy does not collapse, then Theorems 2.8 and 2.9 are of the form "false implies false." However, lowness results provide—right now—hands-on proof of structural simplicity for the sets considered. See Section 2.3 for a discussion of lowness with respect to P-sel.

It is high time to explain what do all these results have to do with selectivity. The answer is that selectivity plays a central role in the proof of both Theorem 2.8 and Theorem 2.9. However, in both cases that role was different, and this difference is the source of the problem described here. In the next few paragraphs we will discuss what this difference is and what challenges it yields.

First, let us concentrate on Theorem 2.8. Hemaspaandra et al. [HNOS96b] used the following technique to obtain the proof. First, they recalled that:

$$\text{NP} \subseteq (\text{NP} \cap \text{coNP})/\text{poly} \implies \text{PH} = \text{NP}^{\text{NP}}$$

([AFK89,Käm91] or alternatively this follows—see [HHN+95]—from the Karp–Lipton Theorem[1]). Then they proved that

$$\text{NPSV-sel} \cap \text{NP} \subseteq (\text{NP} \cap \text{coNP})/\text{poly}.$$

The tempting attack would be to use a technique based on the one employed for Theorem 2.2. However, since NPSV functions may be partial that might seem not to work. Yet, since we are only dealing with selector functions the fact that our functions may be partial is not fatal, since selector functions need to be defined whenever at least one of the inputs is in the set considered. Using this observation and an approach that forces the advice to fork over some very useful membership certificates, one can bring the strategy behind Theorem 2.2's proof successfully to bear even on the case of partial functions.

Let us go back to Theorem 2.8. It is easy to see that if all NP2V functions have NPSV refinements then NP $\subseteq$ NPSV-sel: For any NP set $A$ we can build an NP2V function $f_A$ that takes as its input two strings, nondeterministically guesses one of them that it will act on, nondeterministically guesses a potential certificate that that string is in the set, and on the current path outputs the chosen string if the path's guessed potential certificate is indeed a certificate for the string it was guessed for. Note that each NPSV refinement of $f_A$ is an NPSV-selector for $A$. Thus, by cascading the two above implications, one gets the conclusion of Theorem 2.8.

The first implication is not really related to selectivity, and it can be replaced without affecting the rest of the proof by a similar (stronger) result whenever one becomes available. In fact, we lied above. Hemaspaandra et al. [HNOS96b] actually already drew on such a stronger result, namely, the result of Köbler and Watanabe [KW95,KW98] that if NP is contained in (NP $\cap$ coNP)/poly then the polynomial hierarchy collapses to $\text{ZPP}^{\text{NP}}$. By plugging this result into the above reasoning we get an improved version of Theorem 2.8.

**Theorem 2.10 ([HNOS96b])** *If* NP2V $\subseteq_c$ NPSV *then* PH $= \text{ZPP}^{\text{NP}}$.

Later on Sengupta (see [Cai01]) observed that Hopcroft's proof [Hop81] of the Karp–Lipton Theorem actually shows that NP $\subseteq$ P/poly $\implies$ PH $\subseteq$ $S_2$ (where $S_2$ is the symmetric alternation class of Canetti, Russell, and Sundaram [Can96,RS98]). Since Cai [Cai01] showed that $S_2 \subseteq \text{ZPP}^{\text{NP}}$, Sengupta's result became a stronger Karp–Lipton-style theorem than the Köbler–Watanabe result. Recently, Cai et al. [CCHO05] showed a version of the theorem that is more interesting with respect to our quest, namely, that NP $\subseteq$ (NP $\cap$ coNP)/poly $\implies$ PH $= S_2^{\text{NP} \cap \text{coNP}}$. As noted in [CCHO05], this immediately gives the following theorem.

**Theorem 2.11** *If* NP2V $\subseteq_c$ NPSV *then* PH $= S_2^{\text{NP} \cap \text{coNP}}$.

---

[1]The Karp–Lipton Theorem says that NP $\subseteq$ P/poly $\implies$ NP $= \text{NP}^{\text{NP}}$. This result was first proved by Karp and Lipton in [KL80]. The same paper contained similar results for some other complexity classes, e.g., PSPACE. Since that time many improvements of Karp–Lipton-style results have been obtained, and some of these are mentioned in Section 2.2.

Currently, this is the strongest known collapse of the polynomial hierarchy implied by NP2V having NPSV refinements. So, the sharp progress with respect to Karp–Lipton-style results that has occurred over the years has underpinned a significantly strengthening of Theorem 2.8.

It is natural to ask whether the same strategy can be applied to Theorem 2.9. Unfortunately, that does not seem to be the case. While the proof of Theorem 2.9 is based on an argument that is in some sense a descendant of the argument behind the proof of Theorem 2.2, its nature is different. As a consequence, progress with respect to Karp–Lipton implications buys us nothing with respect to Theorem 2.9 (or at least it seems to buy us nothing).

In the light of the above comments, it is natural to ask if Theorem 2.9 can somehow be improved, and if so then how much? In particular, can it match Theorem 2.11? This question remains unsolved even if we restrict ourselves to NP3V $\subseteq_c$ NP2V. It is an open issue whether NP3V $\subseteq_c$ NP2V $\implies$ PH = $S_2^{NP \cap coNP}$, or even whether NP3V $\subseteq_c$ NP2V $\implies$ PH = $ZPP^{NP}$. The natural indirect way of solving this problem by showing that NP3V $\subseteq_c$ NP2V implies that NP $\subseteq$ (NP $\cap$ coNP)/poly does not seem to work. (However, we do not claim that such a route toward the proof has to be unsuccessful. We only indicate that several people have tried this approach and failed.)

The issues mentioned in this section may also be considered in a slightly different setting. One can, for example, consider refinements of NPMV functions whose number of outputs is not bounded by a constant, but by some function of the length of the input. By a minor abuse of notation, we will denote the subset of NPMV functions with output multiplicity bounded by some function $f$ by NP$f(n)$V. Ogihara [Ogi96] showed the following theorem.

**Theorem 2.12 ([Ogi96])** *Let $\alpha < 1$. If every NPMV function has a NP$(n^\alpha)$V refinement then* PH = $\Sigma_2^p$.

The issues raised in this section with respect to classes NP$k$V can also be asked with respect to the classes NP$(n^\alpha)$V. In particular, we can ask whether Theorem 2.12 can be improved to collapse the polynomial hierarchy more powerfully, e.g., to $ZPP^{NP}$ or $S_2^{NP \cap coNP}$. Another interesting question is how far can one push the $\alpha$ coefficient in Theorem 2.12? On its surface, the approach of Ogihara [Ogi96] seems to die around linear, but is there some way around that?

## 2.3 Question 4: Where does P-sel fall in the extended low hierarchy?

Before describing the main issue of this section, let us present some background regarding the notions of lowness, the low hierarchy, and the extended low hierarchy.

The idea of lowness was brought from recursive function theory to complexity theory by Schöning [Sch83] to explore the internal structure of NP: An NP set $A$ is said to be low for a relativizable complexity class $C$ if and only if it holds that $C^A = C$. In other words, an NP set $A$ is low for $C$ if it gives no extra power to $C$ when used as an oracle. For example, it is known that $NP^{NP \cap coNP} = NP$, and so any member of NP $\cap$ coNP is low for NP. On the other hand, if $A$ is an NP-complete set then $\text{NP} \neq \text{NP}^A$ unless the polynomial hierarchy collapses to NP. The class of sets low for $C$ is denoted as $\text{L}_C$.

The low hierarchy was defined by Schöning ([Sch83], see also [KS85]) as the sequence of sets $\text{L}_{\Sigma_0^p}, \text{L}_{\Sigma_1^p}, \text{L}_{\Sigma_2^p}, \ldots$. Later on intermediate levels, $\text{L}_{\Delta_i^p}$ and $\text{L}_{\Theta_i^p}$,[2] were added by Ko and Schöning [KS85] and Long and Sheu [LS95]. The name low hierarchy is justified by the fact that the following relations hold for every nonnegative value of $k$:

$$\text{L}_{\Sigma_k^p} \subseteq \text{L}_{\Theta_{k+1}^p} \subseteq \text{L}_{\Delta_{k+1}^P} \subseteq \text{L}_{\Sigma_{k+1}^p}.$$

However, the low hierarchy is not known to be a properly infinite hierarchy. If it were then $\text{P} \neq \text{NP}$ would follow. If any NP-complete set belongs to the low hierarchy, the polynomial hierarchy collapses [Sch83].

Lowness results are a type of structural simplicity result. Other types of evidence of structural simplicity include results about containment in nonuniform classes and results implying complexity class collapses. Sometimes a result of one of these types will be so broad as to imply the strongest known results of one of the other types, and sometimes each result type will provide its own separate insight into the simplicity of a class. Regarding the P-selective sets, by Ko's Theorem 2.2, we have the nonuniform simplicity result P-sel $\subseteq$ P/quadratic [Ko83]. This itself implies that P-sel is contained in the low hierarchy, but puts it in a higher level than was obtained in the following direct result of Ko and Schöning [KS85]: P-sel $\cap$ NP $\subseteq \text{L}_{\Sigma_2^p}$. That in turn itself implies that if any NP-complete set belongs to P-sel then the polynomial hierarchy collapses, but the level to which it collapses the polynomial hierarchy is weaker than that to which the hierarchy is collapsed by the following early result of Selman [Sel79].

**Theorem 2.13** *If there exists a* P*-selective* NP*-complete set, then* $\text{P} = \text{NP}$.

Ko and Schöning [KS85], in the same paper in which they proved that P-sel $\cap$ NP is contained in $\text{L}_{\Sigma_2^p}$, also proved lowness upper bounds for P/poly $\cap$ NP and many of its subsets. For many types of structurally simple sets, tight upper and matching relativized lower bounds on their location in the low hierarchy are known ([AH92], see also the surveys [Hem93,Köb95]).

Though lowness results are interesting, the low hierarchy is useless for classifying sets outside of NP, since by definition it applies only to sets in NP. A generalization known as generalized lowness [BBS86] allows lowness to speak meaningfully not just to NP sets but also to sets throughout the polynomial hierarchy.

However, some P-selective sets are not even recursive. Nonetheless, one might hope that the structural simplicity that puts P-sel$\cap$NP into $\text{L}_{\Sigma_2^p}$ might be a reflection of some broader behavior shared by all P-selective sets. To evaluate this intuition, we need a more general version of the low hierarchy, known as the extended low hierarchy.

---

[2]$\Sigma_k^{p,A}$, $\Delta_k^{p,A}$, and $\Pi_k^{p,A}$ are standard levels of the relativized polynomial hierarchy. $\Theta_k^p$ is defined as $\text{P}^{\Sigma_{k-1}^{p,A}[O(\log n)]}$. That is, a $\Theta_k^p$ machine is a deterministic polynomial-time Turing machine that has access to an oracle in $\Sigma_{k-1}^{p,A}$, but may make only $O(\log n)$ queries in total, where $n$ is the length of its input string.

**Definition 2.14** *The levels of the extended low hierarchy are defined by the following:*

1. *For $k \geq 1$, $\mathrm{EL}_{\Sigma_k^p} = \left\{ A \mid \Sigma_k^{p,A} \subseteq \Sigma_{k-1}^{p,A \oplus \mathrm{SAT}} \right\}$.*

2. *For $k \geq 2$, $\mathrm{EL}_{\Delta_k^p} = \left\{ A \mid \Delta_k^{p,A} \subseteq \Sigma_{k-1}^{p,A \oplus \mathrm{SAT}} \right\}$.*

3. *For $k \geq 2$, $\mathrm{EL}_{\Theta_k^p} = \left\{ A \mid \Theta_k^{p,A} \subseteq \Sigma_{k-1}^{p,A \oplus \mathrm{SAT}} \right\}$.*

The extended low hierarchy was defined by Balcázar, Book, and Schöning [BBS86], and the intermediate levels, $\mathrm{EL}_{\Delta_k^p}$ and $\mathrm{EL}_{\Theta_k^p}$, were added by Allender and Hemachandra [AH92] and Long and Sheu [LS95]. As in the case of the low hierarchy, the natural set of containments holds for the extended low hierarchy ($k \geq 1$):

$$\mathrm{EL}_{\Sigma_k^p} \subseteq \mathrm{EL}_{\Theta_{k+1}^p} \subseteq \mathrm{EL}_{\Delta_{k+1}^p} \subseteq \mathrm{EL}_{\Sigma_{k+1}^p}.$$

There are many other analogies between the two hierarchies, but there are also surprising differences. For example, the levels of the low hierarchy are closed downward under many-one reductions, but the same is not true for the classes $\mathrm{EL}_{\Sigma_k^p}$, $k > 1$ [AH92, Ver94]. Another significant difference is that the extended low hierarchy is known to be infinite [SL94], but the low hierarchy is infinite only if $\mathrm{P} \neq \mathrm{NP}$. As a consequence, most lowness lower-bound results are conditional and most extended lowness results are absolute.

Many of the results originally limited to the realm of NP were translated to the world of extended lowness. The paper by Allender and Hemachandra [AH92] can be used as a good starting point if one is interested in this topic. While the primary goal of that article was to establish tight lower bounds on the location of certain classes in the extended low hierarchy, it also contains a summary of previous work, especially regarding upper bounds. The following theorem precisely pinpoints the location of the sparse sets within the extended low hierarchy.

**Theorem 2.15**   1. *[LS95] $\mathrm{SPARSE} \subseteq \mathrm{EL}_{\Theta_3^p}$.*

2. *[AH92] $\mathrm{SPARSE} \not\subseteq \mathrm{EL}_{\Sigma_2^p}$.*

At first it might seem that similar tight bounds should also exist for the P-selective sets. However, obtaining them appears to be surprisingly difficult. The location of the P-selective sets within the extended low hierarchy is the focus of the following results.

**Theorem 2.16**   1. *[ABG90] P-sel $\subseteq \mathrm{EL}_{\Sigma_2^p}$.*

2. *[AH92] There exists a set $A$ such that $\mathrm{P}^A$-sel $\not\subseteq \mathrm{EL}_{\Delta_2^{p,A}}$.*

Of course this is not exactly what we are after, since the lower bound regards a relativized world. A perhaps surprising twist came from the work of Hemaspaandra et al. [HNOS96a], where the following theorem was proven.

**Theorem 2.17 (Implicit in [HNOS96a])** *If $\mathrm{P} = \mathrm{PP}$ then P-sel $\subseteq \mathrm{EL}_{\Sigma_1^p}$.*

Thus, under an implausible, but nonetheless possible, complexity-theoretic assumption it is the case that the P-selective sets are in $\mathrm{EL}_{\Sigma_1^p}$. Of course, a proof that this is not the case would be of enormous value since it would imply $\mathrm{P} \neq \mathrm{PP}$. Theorem 2.17 is also surprising because the connection between P-sel and PP is not apparent at first. However, the surprise is slightly reduced if one recalls that the first result saying that P-selective sets have linear advice, Theorem 2.3, used PP as the advice interpreter.

Before we discuss the proof of Theorem 2.17 in a little more detail, let us give the definition of the standard left cut of a real number.

**Definition 2.18** *Let $r$ be any real number such that $0 \leq r < 1$. The standard left cut of $r$ is defined as $L(r) = \left\{ b_1 b_2 b_3 \ldots b_z \mid (\forall j : 1 \leq j \leq z)[b_j \in \{0,1\}] \wedge r > \sum_{i=1}^{z} \frac{b_i}{2^i} \right\}$.*

It is not difficult to notice that every standard left cut of some real number $r$ is P-selective: The selector function simply interprets its inputs as numbers (in the obvious way suggested by the above definition) and chooses the smaller one. Also, standard left cuts require only a linear amount of advice since at every length $n$ it is sufficient to provide the first $n$ bits of $r$ and one more bit saying whether $r$ is greater than the number defined by these first $n$ bits.

Theorem 2.17 can be divided into three parts that together give the same conclusion:

1. If $\mathrm{P} = \mathrm{PP}$ then for every infinite P-selective set $A$ there exists a standard left cut $L(r)$ such that $A$ is many-one equivalent to $L(r)$ [HNOS96a].

2. Every standard left cut is Turing equivalent to some tally language [Sel81].

3. Every set Turing equivalent to a tally language is in $\mathrm{EL}_{\Sigma_1^p}$ (Theorem 3.1 of [BB86], plus the observation two paragraphs before Theorem 4.2 of [BB86]).

As opposed to Theorem 2.3, Theorem 2.17—or rather its first part—does not seem to have a simple refinement that uses a class smaller than PP. The proof given by Hemaspaandra et al. strongly relies on the ability to count the number of strings that lose to a given input $x$ in the tournament induced by the P-selective set considered.

As a consequence of the above discussion several open questions arise. Regarding the area opened by Theorem 2.17: Is it possible to prove that P-sel $\subseteq \mathrm{EL}_{\Sigma_1^p}$ using a weaker assumption than $\mathrm{P} = \mathrm{PP}$? And are there any nice, complexity-theoretic consequences that follow from the assumption that P-sel $\subseteq \mathrm{EL}_{\Sigma_1^p}$? A most welcome result would be to give matching necessary and sufficient conditions for P-sel $\subseteq \mathrm{EL}_{\Sigma_1^p}$. Such conditions would be particularly exciting if they could be expressed in terms of complexity classes collapses. Another interesting research direction would be to attempt to prove (or disprove, though that would imply $\mathrm{P} \neq \mathrm{PP}$) that every P-selective set is many-one equivalent to some standard left cut.

## 2.4 Question 5: Can an NP-hard set be P-selective?

The previous section dealt with the structural simplicity of P-selective sets in terms of lowness theory. Now we will discuss the same matter in a somewhat different setting, namely, we will be interested whether P-selective sets, given some reduction type, can be hard for certain complexity classes. In particular, we will be interested in P-selective sets being hard for NP with respect to truth-table reductions.

To facilitate the discussion, we need to provide some notation. Let $\leq_b^a$ be some already defined reduction type, and let $\mathcal{C}$ be a complexity class. By $\mathrm{R}_b^a(\mathcal{C})$ we mean the set of all languages $\leq_b^a$-reducible to sets in $\mathcal{C}$. More formally,

$$\mathrm{R}_b^a(\mathcal{C}) = \{A \mid (\exists B \in \mathcal{C})[A \leq_b^a B]\}.$$

One of the characterizations of the class P/poly is the following well-known theorem (whose different equalities come from different papers).

**Theorem 2.19** $\mathrm{R}_T^p(\text{P-sel}) = \mathrm{R}_T^p(\text{SPARSE}) = \mathrm{R}_{tt}^p(\text{TALLY}) = \text{P/poly}$.

The simplest case of our problem arises when working with many-one reductions. By Theorem 2.13, if any NP-complete set is P-selective then P = NP.

The case of Turing reductions is also relatively well studied. By Theorem 2.19, the class of languages Turing reducible to P-selective sets is exactly the class P/poly, and so the line of research regarding Karp–Lipton-style theorems handles this case. We will not repeat the discussion of Karp–Lipton results of Section 2.2, but instead we will just mention currently the best known collapse.

**Theorem 2.20 ([Ko83] in light of [Cai01])** *If* $\text{NP} \subseteq \mathrm{R}_T^p(\text{P-sel})$ *then* $\text{PH} = \text{S}_2$.

However, the situation is not that simple in the case of intermediate reduction types. For the rest of this section we will concentrate on truth-table reductions:

**Definition 2.21 ([LLS75])** *A language $A$ is truth-table reducible to language $B$, denoted $A \leq_{tt}^p B$, if there exist two polynomial-time functions $f$ and $g$ such that for each string $x$:*

- *$f(x)$ is a list of strings $\langle x_1, \ldots, x_m \rangle$, and*

- *$g(\langle x_1, \chi_B(x_1), x_2, \chi_B(x_2), \ldots, \chi_B(x_m) \rangle) = 1$ if and only if $x \in A$.*

*Let $k$ be some positive integer constant. If on every input $x$ the function $f$ computes a list of $k$ strings then we say that $A$ is $k$-truth-table reducible to $B$, denoted $A \leq_{k\text{-}tt}^p B$. If $(\exists k)[A \leq_{k\text{-}tt}^p B]$, then we say that $A$ is bounded-truth-table reducible to $B$.*

One can look at a truth-table reduction as if it were a Turing reduction that asks all of its queries at the same time. In other words, the queries cannot be adaptive.

It would be desirable to have a result similar to Theorem 2.13 for truth-table reductions, i.e., we would like to be able to say that if an NP truth-table-hard set were P-selective then P would equal NP. Unfortunately, currently only weaker conclusions seem to follow from the desired hypothesis, and to get the desired conclusion we seem to need a stronger hypothesis. This is especially disturbing since similar results are known to hold for UP and PSPACE. Let us briefly discuss how they were obtained.

The key method used to show that P-selective sets cannot be truth-table hard for certain complexity classes (unless some unlikely collapses occur) is Toda's minimum path technique. Using this approach, Toda [Tod91] showed that UP and PSPACE cannot have P-selective truth-table-hard sets unless P = UP and P = PSPACE.

**Theorem 2.22 ([Tod91])**  1. P = UP *if and only if there is a* UP-$\leq_{tt}^p$ *-hard* P*-selective set.*

2. P = NP *if and only if there is a* $\Delta_2^p$-$\leq_{tt}^p$ *-hard* P*-selective set.*

3. P = PSPACE *if and only if there is a* PSPACE-$\leq_{tt}^p$ *-hard* P*-selective set.*

The first two parts of the above theorem are proved using the minimum path technique, and the third one is a consequence of the second part if one invokes the fact[3] that PSPACE $\subseteq$ P/poly $\implies$ PSPACE = $\Sigma_2^p$ [KL80] and Theorem 2.19.

The minimum path technique works as follows. Let us fix some NP-type machine $M$ whose language we will attempt to decide in polynomial time. We are given an input string $x$, and our goal is to figure out the lexicographically minimum accepting computation path of $M$ on $x$, provided one exists. We assume that at every step $M$ has exactly two nondeterministic choices labeled 0 and 1 and that a polynomial $p$ exists such that every computation path of $M$ on $x$ has length $p(|x|)$. Each NP machine can easily be converted to a machine meeting these requirements and accepting the same language as the original machine.

The minimum accepting path is constructed using the "minimum-path" language, that is, the language of all strings of the form $\langle x, j \rangle$ such that $x \in L(M)$ and $M$'s lexicographically minimal path makes a nondeterministic decision 0 at its $j$th step (provided it has at least $j$ steps). Of course, if one could answer queries to this language in polynomial time then one could easily construct the minimum accepting path. However, deciding the "minimum-path" language in polynomial time seems to be more difficult than the original task. Instead of answering such queries one at a time, we shall use a clever trick that will allow us to answer them (well—not answer them, but learn something about the plausible answers) in bulk, provided the "minimum-path" language is truth-table reducible to some P-selective set.

So, let us assume that the "minimum-path" language is truth-table reducible to some P-selective set $A$, and let us say that this reduction is computed by functions $f$ and $g$, as in Definition 2.21. For every $1 \leq j \leq p(|x|)$ compute the list of strings $f(\langle x, j \rangle)$, and let $y_1, \ldots, y_{q(|x|)}$ be the union of these lists. ($q$ is bounded by some polynomial.) If we tried all $2^{q(|x|)}$ possible combinations of $y_i$'s being or not being members of $A$ then we would certainly find the minimum accepting path if one existed, but of course, it would take exponentially many steps. Fortunately, by the Toda Ordering Lemma [Tod91], it is enough to only try $q(x) + 1$ combinations.

---

[3]Note that this again is a Karp–Lipton-style result, of which a stronger version is known but is not needed here.

**Theorem 2.23 ([Tod91])** *There exists a polynomial-time function h such that:*

1. *On input $\langle x_1, \ldots, x_k \rangle$, h outputs a permutation of its input strings, $\langle x_{\sigma(1)}, \ldots, x_{\sigma(k)} \rangle$.*

2. *$\{x_1, \ldots, x_k\} \cap A = \{x_{\sigma(1)}, \ldots, x_{\sigma(i)}\}$ for some $0 \leq i \leq k$*

The reason the Toda Ordering Lemma holds is we can use our P-selector function $f$ to make $\sigma$ be such that $(\forall j : 1 \leq j \leq i-1)[f(x_{\sigma(j)}, x_{\sigma(j+1)}) = x_{\sigma(j)}]$. After applying the Toda Ordering Lemma to $y_1, \ldots, y_{q(|x|)}$ it is enough to check the following combinations: no $y_i$ is in $A$, only $y_{\sigma(1)}$ is in $A$, only $y_{\sigma(1)}$ and $y_{\sigma(2)}$ are in $A$, and so on. Clearly, if the minimum accepting path exists then by inspecting separately each of the above answers to the "minimum-path" language we will be able to find some accepting path, since the correct one of the above will actually do the right thing.

Unfortunately, the minimum path technique does not seem to be capable of yielding "P = NP if and only if there is a P-selective set that is truth-table hard for NP," but only the weaker second part of Theorem 2.22. The reason is that the "minimum-path" language for a generic NP machine belongs to $\Delta_2^p$. (On the other hand, the "minimum path" language for a "UP machine" is itself in UP.) The following theorem lists a few of the consequences that are known to follow from the assumption that there is a P-selective set that is truth-table hard for NP (see also the work of Sivakumar [Siv99] regarding working with a weaker hypothesis).

**Theorem 2.24** *If there is a P-selective set that is truth-table hard for NP then*

1. *[Tod91] P = FewP, and*

2. *[Bei88,Tod91] R = NP.*

Since every truth-table reduction implies a Turing reduction, by Theorem 2.20 we could add PH = $S_2$ to the above list. Some further consequences can also be added, see, e.g., the work of Naik and Selman [NS99] or the treatment in [HT03].

A natural direction is to seek how much more do we need to assume to obtain P = NP. In particular, is Theorem 2.13 the best we can do? The following theorem shows that the answer is "no."

**Theorem 2.25** *P = NP if and only if there is a P-selective set that is bounded-truth-table hard for NP.*

Theorem 2.25 was proved independently by Ogihara [Ogi95], by Agrawal and Arvind [AA96], and by Beigel, Kummer, and Stephan [BKS95]. The fact that so many papers obtained the same result is especially interesting as each of these papers is actually dealing with somewhat different issues than the others. Ogihara introduced a generalization of P-sel, the polynomial-time membership comparable sets, Agrawal and Arvind were interested in quasi-linear truth-table reductions, and Beigel, Kummer, and Stephan explored p-superterseness of NP. Of course, there were also some unifying themes among these papers, such as selectivity, self-reducibility, and truth-table reductions. Even the claim of Theorem 2.25 is not the last word. The behavior holds even up to $n^{1-\epsilon}$-truth-table reductions [Ogi95].

It is worth mentioning that the issues of this section can also be studied with respect to NPSV-selective sets (see [HNOS96b,HHN$^+$95]). In fact, this is closely related to Question 3, due to the fact that SAT is NPSV-selective if and only if NPMV $\subseteq_c$ NPSV ([HNOS96b], see also [Sel94]). So Theorem 2.11 can be restated as follows.

**Theorem 2.26** *If* SAT *is* NPSV*-selective then* PH $= S_2^{NP \cap coNP}$.

## 3 Conclusions

In this paper we have presented some open questions in the theory of P-selective and NPSV-selective sets (in the theory of semifeasible computation). While we have chosen the problems to present according to our taste, we believe that they are important and attractive. The amount of attention these problems received in the recent years support our choice. Yet, of course there are also other problems, and we certainly do not claim that the ones we presented are more important. We do hope that this paper will attract even more researchers to invest their time in the theory of semifeasible computation.

**Acknowledgment** We thank Alan L. Selman for helpful comments on the background of function refinement.